\documentstyle[prd,aps,preprint,tighten,epsfig]{revtex}

\begin{document}

\draft

\title{Wolfenstein-like Parametrization of the Neutrino Mixing Matrix}
\author{\bf Zhi-zhong Xing}
\address{CCAST (World Laboratory), P.O. Box 8730, Beijing 100080, China \\
and Institute of High Energy Physics, Chinese Academy of Sciences, \\
P.O. Box 918 (4), Beijing 100039, China 
\footnote{Mailing address} \\
({\it Electronic address: xingzz@mail.ihep.ac.cn}) }
\maketitle

\begin{abstract}
We show that the $3\times 3$ lepton flavor mixing matrix $V$ can 
be expanded in powers of a Wolfenstein-like parameter 
$\Lambda \equiv |V_{\mu 3}| \sim 1/\sqrt{2}~$, which measures the strength 
of flavor conversion in atmospheric neutrino oscillations. The term of 
${\cal O}(\Lambda^2)$ is associated with the large mixing angle
in solar neutrino oscillations. The Dirac phase of CP violation enters
at or below ${\cal O}(\Lambda^8)$, and the Majorana phases of CP violation
are not subject to the $\Lambda$-expansion. Terrestrial matter effects on 
this new parametrization in realistic long-baseline neutrino oscillation 
experiments are briefly discussed. Some comments are also
given on the possible relation between $\Lambda$ and a relatively weak 
hierarchy of neutrino masses.
\end{abstract}


\newpage

The KamLAND neutrino experiment \cite{Kam} has recently confirmed the 
large-mixing-angle (LMA) Mikheyev-Smirnov-Wolfenstein (MSW) 
solution \cite{MSW} to the solar neutrino problem. Meanwhile, the 
K2K long-baseline neutrino experiment \cite{K2K} has unambiguously 
observed a reduction of $\nu_\mu$ flux and a distortion of the energy 
spectrum. These new measurements, together with the robust SNO 
evidence \cite{SNO} for the flavor conversion of solar $\nu_e$ neutrinos 
and the compelling Super-Kamiokande evidecne \cite{SK} for the deficit 
of atmospheric $\nu_\mu$ neutrinos, convinces us that the hypothesis 
of neutrino oscillations is indeed correct! We are then led to the 
exciting conclusion that neutrinos do have masses and lepton flavor
mixing does exist.

A global analysis of today's solar neutrino data \cite{Fogli} indicates
that the maximal mixing is strongly disfavored for the LMA solution. 
The mixing factor of atmospheric neutrino oscillations is found to be 
almost maximal \cite{ATM}, on the other hand. Taking account of the 
KamLAND, K2K, SNO, Super-Kamiokande and CHOOZ results \cite{CHOOZ}, 
we expect that the $3\times 3$ lepton flavor mixing 
matrix $V$ is typically of a constant pattern \cite{Giunti} in the 
leading-order approximation:
\begin{equation}
V \; =\; \left ( \matrix{
\displaystyle\frac{\sqrt{3}}{2}	& \displaystyle\frac{1}{2}	& 0 \cr\cr
-\displaystyle\frac{\sqrt{2}}{4}	
& ~~~ \displaystyle\frac{\sqrt{6}}{4} ~~~	
& \displaystyle\frac{\sqrt{2}}{2} \cr\cr
\displaystyle\frac{\sqrt{2}}{4}		& -\displaystyle\frac{\sqrt{6}}{4}
& \displaystyle\frac{\sqrt{2}}{2} \cr}
\right ) \; .
\end{equation} 
Namely, 
\begin{equation}
\left \{ \theta_{\rm sun}, \theta_{\rm atm}, \theta_{\rm chz} \right \} = 
\left \{ 30^\circ, 45^\circ, 0^\circ \right \} \; ,
\end{equation}
or $\sin^2 2\theta_{\rm sun} = 3/4$, 
$\sin^2 2\theta_{\rm atm} = 1$ and $\sin^2 2\theta_{\rm chz} = 0$. Note
that $|V_{e2}| = |V_{\mu 3}|^2$ holds in this simplified neutrino mixing 
pattern. It implies that an expansion of $V$ in terms of $V_{\mu 3}$ is
actually possible. Note also that one may introduce a small perturbation 
into $V$, such that 
\begin{enumerate}
\item	$\theta_{\rm sun}$ gets closer to its best fit value 
($\theta_{\rm sun} \sim 32^\circ$ \cite{Fogli});
\item	$\sin^2 2\theta_{\rm atm}$ deviates slightly from unity
($\sin^2 2\theta_{\rm atm} > 0.92$ \cite{ATM}); 
\item	$|V_{e3}| = \sin\theta_{\rm chz} \neq 0$ appears
($\sin^2 2\theta_{\rm chz} < 0.1$ \cite{CHOOZ});
\item	a Dirac phase of CP violation can naturally be included into 
$V$ \cite{FX01}.
\end{enumerate}
Motivated by these observations, we proceed to propose a new parametrization
of the lepton flavor mixing matrix $V$, in which all matrix elements are 
expanded in powers of a parameter $\Lambda \equiv |V_{\mu 3}| \sim 1/\sqrt{2} ~$. 
Clearly $\Lambda$ can be regarded as the leptonic analog of the well-known
Wolfenstein parameter $\lambda \equiv |V_{us}| \approx 0.22$ for quark flavor 
mixing \cite{Wol}.

The first step is to parametrize $V$ in the limit of $V_{e3} =0$. We obtain
\begin{equation}
V \; =\; \left ( \matrix{
\sqrt{1-A^2 \Lambda^4}  & A \Lambda^2  & 0 \cr\cr
-A \Lambda^2 \sqrt{1-\Lambda^2}  & ~ \sqrt{\left (1-\Lambda^2 \right )
\left (1-A^2\Lambda^4 \right )} ~  & \Lambda \cr\cr
A\Lambda^3  & -\Lambda \sqrt{1-A^2\Lambda^4}  & \sqrt{1-\Lambda^2} \cr}
\right ) \; ,
\end{equation}
where $A$ is a positive coefficient of ${\cal O}(1)$. If $A =1$ and
$\Lambda =1/\sqrt{2}$ are taken, the constant pattern of $V$ in Eq. (1) 
can straightforwardly be reproduced from Eq. (3). It is obvious that 
$\Lambda$ measures the strength of flavor mixing in atmospheric neutrino
oscillations, while $A\Lambda^2$ characterizes the magnitude of flavor
mixing in solar neutrino oscillations.

The second step is to introduce small corrections to $V$ in Eq. (3), such
that $|V_{e3}| \neq 0$ appears. Because $|V_{e3}| < 0.16$ is 
required \cite{CHOOZ}, we may take 
$|V_{e3}| \sim {\cal O}(\Lambda^8) \sim 0.06$ as a typical possibility for 
$\Lambda \sim 1/\sqrt{2} ~$. Smaller values of $|V_{e3}|$ are certainly 
allowed. In a number of phenomenological models for lepton flavor 
mixing \cite{FX99}, however, $|V_{e3}| \sim \sqrt{m_e/m_\mu} \sim 0.07$ 
is naturally predicted. Hence ${\cal O}(\Lambda^8)$ could be the plausible 
order of $|V_{e3}|$. Let us fix the matrix elements $V_{e2}$, $V_{e3}$ and
$V_{\mu 3}$ by use of four independent parameters: 
\begin{equation}
V_{\mu 3} = \Lambda \; , ~~
V_{e2} = A\Lambda^2 \; , ~~
V_{e3} = B \Lambda^8 e^{-i\delta} \; ,
\end{equation}
where $B$ is of ${\cal O}(1)$ or smaller, and $\delta$ denotes the Dirac phase 
of leptonic CP violation. Given Eq. (4), one may make use of the unitarity 
of $V$ to work out exact analytical expressions for the other six matrix 
elements. The relevant results are quite complicated and will be presented
elsewhere. We find that it is more instructive to approximate $V$ as
\begin{equation}
V \; =\; \left ( \matrix{
\sqrt{1-A^2 \Lambda^4}  & A \Lambda^2  & B\Lambda^8 e^{-i\delta} \cr\cr
-A \Lambda^2 \sqrt{1-\Lambda^2}  & \sqrt{\left (1-\Lambda^2 \right )
\left (1-A^2\Lambda^4 \right )}  & \Lambda \cr\cr
\Lambda^3 \left [A - B\Lambda^5 \sqrt{\left (1-\Lambda^2 \right )
\left (1-A^2\Lambda^4 \right )} ~ e^{i\delta} \right ]
& -\Lambda \sqrt{1-A^2\Lambda^4}  & \sqrt{1-\Lambda^2} \cr}
\right ) \; .
\end{equation}
In this approximation, the unitary normalization relations of $V$ keep
valid to ${\cal O}(\Lambda^{11}) \sim 2\%$. Hence Eq. (5) is sufficiently
accurate to describe lepton flavor mixing, not only in solar and atmospheric 
neutrino oscillations, but also in some of the proposed long-baseline 
neutrino oscillation experiments where leptonic CP violation is not 
concerned \cite{LBL}. As the unitary orthogonality relations of $V$
in the above approximation are valid to ${\cal O}(\Lambda^8) \sim 6\%$,
the leptonic unitarity triangles can also be described by 
Eq. (5) to a reasonably good degree of accuracy. A rephasing-invariant
measure of leptonic CP violation is the well-known Jarlskog parameter 
$J$ \cite{J}, whose magnitude must be proportional to $\sin\delta$ in
our parametrization. The explicit expression of $J$ reads
\begin{equation}
{\cal J} \; =\; AB\Lambda^{11} \sqrt{\left (1-\Lambda^2 \right )
\left (1-A^2\Lambda^4 \right )} ~ \sin\delta \; .
\end{equation}
Note again that $\Lambda^{11} \sim 0.02$ for $\Lambda \sim 1/\sqrt{2} ~$. Thus
$|\cal J|$ may be at the percent level, if $\delta \sim \pm 90^\circ$ holds.
No matter whether neutrinos are Dirac or Majorana particles, the strength 
of CP and T violation in normal neutrino-neutrino and antineutrino-antineutrino 
oscillations is always governed by $\cal J$ or by the Dirac phase 
$\delta$ \cite{Barger}. Furthermore, the off-diagonal asymmetries of the
lepton flavor mixing matrix $V$ \cite{Xing96} read as
\begin{eqnarray}
{\cal A}_{\rm L} & \equiv & |V_{e2}|^2 - |V_{\mu 1}|^2 = 
|V_{\mu 3}|^2 - |V_{\tau 2}|^2 = |V_{\tau 1}|^2 - |V_{e3}|^2
\nonumber \\
& = & A^2 \Lambda^6 \; ,
\nonumber \\
{\cal A}_{\rm R} & \equiv & |V_{e2}|^2 - |V_{\mu 3}|^2 = 
|V_{\mu 1}|^2 - |V_{\tau 2}|^2 = |V_{\tau 3}|^2 - |V_{e1}|^2
\nonumber \\
& = & \Lambda^2 \left (A^2 \Lambda^2 -1 \right ) \; .
\end{eqnarray}
We see that ${\cal A}_{\rm L} >0$ holds definitely. In comparison, the sign
of ${\cal A}_{\rm R}$ cannot be fixed from the present
experimental data. It is actually possible to obtain ${\cal A}_{\rm R} =0$,
when $A^2\Lambda^2 =1$ is satisfied. In this interesting case, the lepton 
flavor mixing matrix $V$ is exactly symmetric about its 
$V_{e3}$-$V_{\mu 2}$-$V_{\tau 1}$ axis \cite{Xing02a}.

The final step is to incorporate $V$ with two Majorana phases of CP violation,
provided neutrinos are Majorana particles. To do so, we simply multiply $V$ 
on its right-hand side with a pure phase matrix; i.e., 
\begin{equation}
V \; \Longrightarrow \; V P \; , ~~~
P \; =\; \left ( \matrix{
e^{i\rho}  & 0  & 0 \cr\cr
0  & e^{i\sigma}  & 0 \cr\cr
0  & 0  & e^{i\delta} \cr} \right ) \; ,
\end{equation}
in which $\rho$ and $\sigma$ are the Majorana-type CP-violating 
phases \cite{FX01}. The phase convention of $P$ chosen in Eq. (8) 
is to make the CP-violating phase $\delta$ not to manifest 
itself in the effective mass term of the neutrinoless double beta decay:
\begin{equation}
\langle m\rangle_{ee} \; = \; \left | m_1 \left (1 - A^2\Lambda^4
\right ) e^{2i\rho} + m_2 A^2 \Lambda^4 e^{2i\sigma} +
m_3 B \Lambda^{16} \right | \; ,
\end{equation}
where $m_i$ (for $i=1,2,3$) are physical neutrino masses. This result
can somehow get simplified, if a specific pattern of the neutrino mass
spectrum is assumed. The present experimental upper bound is 
$\langle m\rangle_{ee} < 0.35 ~ {\rm eV}$ 
(at the $90\%$ confidence level \cite{HM}), from which no constraint 
on $\rho$ and $\sigma$ can be got.  

Now let us establish the direct relations between 
($\Lambda$, $A$, $B$) and ($\theta_{\rm atm}$, $\theta_{\rm sun}$,
$\theta_{\rm chz}$). With the help of Eq. (4) and 
\begin{eqnarray}
|V_{e2}|^2 & = & \frac{\cos^2\theta_{\rm chz}}{2}
- \frac{\sqrt{\cos^4\theta_{\rm chz} - \sin^2 2\theta_{\rm sun}}}{2} \; ,
\nonumber \\
|V_{e3}|^2 & = & \sin^2 \theta_{\rm chz} \; ,
\nonumber \\
|V_{\mu 3}|^2 & = & \sin^2 \theta_{\rm atm} \; ,
\end{eqnarray}
which have been obtained in Ref. \cite{Xing02b}, we arrive at
\begin{eqnarray}
\Lambda & = & \sin\theta_{\rm atm} \; ,
\nonumber \\
A & = & \frac{\sqrt{\cos^2\theta_{\rm chz} - \sqrt{\cos^4\theta_{\rm chz}
- \sin^2 2\theta_{\rm sun}}}}{\sqrt{2} \sin^2\theta_{\rm atm}} \; ,
\nonumber \\
B & = & \frac{\sin\theta_{\rm chz}}{\sin^8\theta_{\rm atm}} \; .
\end{eqnarray}
Once the mixing angles $\theta_{\rm atm}$, $\theta_{\rm sun}$ and 
$\theta_{\rm chz}$ are precisely measured, we may use Eq. (11) to
determine the magnitudes of $\Lambda$, $A$ and $B$. For the purpose of
illustration, we typically take 
$0.25 \leq \sin^2\theta_{\rm sun} \leq 0.40$ \cite{Fogli},
$\sin^2 2\theta_{\rm atm} > 0.92$ \cite{ATM} and 
$\sin^2 2\theta_{\rm chz} < 0.1$ \cite{CHOOZ} to calculate the 
allowed regions of $\Lambda$, $A$ and $B$. Then we obtain
$0.6 \leq \Lambda \leq 0.8$ straightfowardly. The numerical results for 
$A$ and $B$ are presented in Fig. 1, from which $0.8 \leq A \leq 1.94$ and
$0 \leq B \leq 9.5$ can directly be read off. 

It is also worthwhile to connect ($\Lambda$, $A$, $B$) to 
($\theta_{12}$, $\theta_{23}$, $\theta_{13}$), which are three mixing 
angles of the ``standard parametrization'' of $V$ \cite{PDG}. We find 
\begin{eqnarray}
\sin\theta_{12} & = & \frac{A\Lambda^2}{\sqrt{1-B^2\Lambda^{16}}} 
\; \approx \; A\Lambda^2 \; ,
\nonumber \\
\sin\theta_{23} & = & \frac{\Lambda}{\sqrt{1-B^2\Lambda^{16}}} 
\; \approx \; \Lambda \; ,
\nonumber \\
\sin\theta_{13} & = & B\Lambda^8 \; .
\end{eqnarray}
In addition, the Dirac phase of CP violation in the standard 
parametrization is exactly equal to $\delta$ defined in the present
Wolfenstein-like parametrization. 

An interesting point is that the {\it effective} lepton flavor mixing
matrix in matter, which is denoted as $\tilde{V}$ \cite{Xing00}, can similarly be 
parametrized in terms of four matter-corrected parameters $\tilde{\Lambda}$,
$\tilde{A}$, $\tilde{B}$ and $\tilde{\delta}$:
\begin{equation}
\tilde{V} \; =\; \left ( \matrix{
\sqrt{1-\tilde{A}^2 \tilde{\Lambda}^4}  & 
\tilde{A} \tilde{\Lambda}^2  & 
\tilde{B}\tilde{\Lambda}^8 e^{-i\tilde{\delta}} \cr\cr
-\tilde{A} \tilde{\Lambda}^2 \sqrt{1-\tilde{\Lambda}^2}  & 
\sqrt{(1-\tilde{\Lambda}^2 ) (1-\tilde{A}^2\tilde{\Lambda}^4 )}  & 
\tilde{\Lambda} \cr\cr
\tilde{\Lambda}^3 \left [\tilde{A} - \tilde{B}\tilde{\Lambda}^5 
\sqrt{(1-\tilde{\Lambda}^2 )
(1-\tilde{A}^2\tilde{\Lambda}^4 )} ~ e^{i\tilde{\delta}} \right ] & 
-\tilde{\Lambda} \sqrt{1-\tilde{A}^2\tilde{\Lambda}^4}  & 
\sqrt{1-\tilde{\Lambda}^2} \cr}
\right ) \; .
\end{equation}
Clearly there exist the same relations as those given in
Eq. (12) between the {\it effective} mixing angles of $\tilde{V}$
(i.e., $\tilde{\theta}_{12}$, $\tilde{\theta}_{23}$ and 
$\tilde{\theta}_{13}$) and the corresponding new parameters 
($\tilde{\Lambda}$, $\tilde{A}$ and $\tilde{B}$). It has been shown that
$\sin\tilde{\theta}_{23} \approx \sin\theta_{23}$ and 
$\sin\tilde{\delta} \approx \sin\delta$ hold to leading order for a variety 
of terrestrial long-baseline neutrino oscillation experiments \cite{Xing01}. 
Therefore, we have
\begin{equation}
\tilde{\Lambda} \; \approx \; \Lambda \; , ~~~~~~~
\tilde{\delta} \; \approx \; \delta \; .
\end{equation}
This result implies that $\Lambda$ and $\delta$ are essentially
stable against terrestrial matter effects. Hence the expansion of 
$\tilde{V}$ in powers of $\tilde{\Lambda} \approx \Lambda$ makes sense.
Only $A$ and $B$ in $V$ are sensitive to the matter-induced corrections.
Because of $\tilde{A} \propto \sin\tilde{\theta}_{12}$ and
$\tilde{B} \propto \sin\tilde{\theta}_{13}$, two remarkable conclusions can 
be drawn from Ref. \cite{Xing01} for our new parameters:
(a) $\tilde{A}/A $ is suppressed up to the
order $\Delta m^2_{\rm sun}/\Delta m^2_{\rm atm}$; and 
(b) $\tilde{B}/B$ may have the resonant behavior similar to the
two-neutrino MSW resonance \cite{MSW}.

Finally we give some speculation on the physical meaning of $\Lambda$.
It is well known that the Wolfenstein parameter $\lambda \approx 0.22$ can 
be related to the ratios of quark masses in the Fritzsch ansatz of
quark mass matrices \cite{F78} or its modified versions \cite{FX95}:
\begin{equation}
\lambda \; \approx \; \left |\sqrt{\frac{m_u}{m_c}} - e^{i\phi_\lambda}
\sqrt{\frac{m_d}{m_s}} \right | \; ,
\end{equation}
where $\phi_\lambda$ denotes the phase difference between the (1,2)
elements of up- and down-type quark mass matrices. Confronting 
Eq. (15) with current experimental data on the Cabibbo angle and
quark masses leads to $\phi_\lambda \sim \pm 90^\circ$ \cite{FX95}. 
Such a result for $\phi_\lambda$ is also consistent with the large 
CP-violating effect observed in 
$B^0_d$ vs $\bar{B}^0_d \rightarrow J/\psi K_{\rm S}$ decays at KEK and 
SLAC $B$-meson factories \cite{B}. Eq. (15) indicates that the smallness 
of $\lambda$ is a natural consequence of the strong quark mass 
hierarchy. Could the largeness of $\Lambda$ be attributed to a relatively
weak hierarchy of three neutrino masses? The answer is indeed affirmative in 
the Fritzsch texture of lepton mass matrices, which coincides with current 
experimental data on neutrino oscillations if the masses of three neutrinos 
perform a normal but weak hierarchy (typically, $m_1$ : $m_2$ : $m_3$ 
$\approx$ 1 : 3 : 10) \cite{Xing02c}. In this phenomenological model,
we approximately obtain  
\begin{equation}
\Lambda \; \approx \; \left |\sqrt{\frac{m_2}{m_3}} - e^{i\phi_\Lambda}
\sqrt{\frac{m_\mu}{m_\tau}} \right | \; ,
\end{equation}
where $\phi_\Lambda$ denotes the phase difference between the (2,3)
elements of charged lepton and neutrino mass matrices. We find that
$\phi_\Lambda \sim \pm 180^\circ$ is practically favored \cite{Xing02c},
in order to obtain a sufficiently large $\Lambda$. To illustrate, we 
typically take $m_2/m_3 \sim 0.3$ as well as $m_\mu/m_\tau \approx 0.06$ \cite{PDG}. 
Then we arrive at $\Lambda \sim 0.8$, a result compatible with our
empirical expectation for the order of $\Lambda$
\footnote{Assuming a somehow stronger mass hierarchy for three neutrinos,
Kaus and Meshkov \cite{KM} have proposed a different expansion of
the neutrino mixing matrix in terms of $\Lambda = \sqrt{m_2/m_3} = 
(\Delta m^2_{\rm sun}/\Delta m^2_{\rm atm})^{1/4} \sim 0.37$. This
parameter is associated with $V_{e2}$ instead of $V_{\mu 3}$, therefore
it is sensitive to the matter effect. In contrast, our parametrization
does not rely on the assumption of neutrino mass hierarchy, and its 
expansion parameter is insensitive to the matter-induced corrections.}.

In summary, we have shown that the $3\times 3$ lepton flavor mixing matrix
can actually be expanded in terms of a Wolfenstein-like parameter
$\Lambda \sim 1/\sqrt{2} ~$. This parameter measures the strength of 
flavor mixing in atmospheric neutrino oscillations, thus it is insensitive
to the matter effect. In our new parametrization, the term of
${\cal O}(\Lambda^2)$ is associated with the flavor mixing angle of
solar neutrino oscillations. The Dirac-type CP-violating phase enters 
at or below the level of ${\cal O}(\Lambda^8)$, while the 
Majorana-type CP-violating phases are not subject to the 
$\Lambda$-expansion. Direct relations between the parameters in this 
Wolfenstein-like representation and those in the standard representation
have been established. We expect that such a new description of lepton 
flavor mixing can be very useful in phenomenology of neutrino physics.

Relating our new parametrization to the models of lepton mass matrices 
is not the subject of this short note. Nevertheless, we have taken the
Fritzsch ansatz for example to give a simple interpretation of the 
Wolfenstein parameter $\lambda$ in the quark sector and its analog $\Lambda$ 
in the lepton sector. We find that their different magnitudes reflect 
different mass hierarchies of quarks and leptons. This observation is very
suggestive, although it is quite preliminary. Further attempts are
therefore desirable, towards deeper understanding of both similarities
and differences between lepton and quark mass spectra and flavor mixing
schemes.

\vspace{0.3cm}

The author likes to thank IPPP in University of Durham, where the
paper was written, for its warm hospitality and stimulating atmosphere.
This work was supported in part by National Natural Science Foundation
of China.

\newpage

\begin{figure}[t]
\vspace{-2.7cm}
\epsfig{file=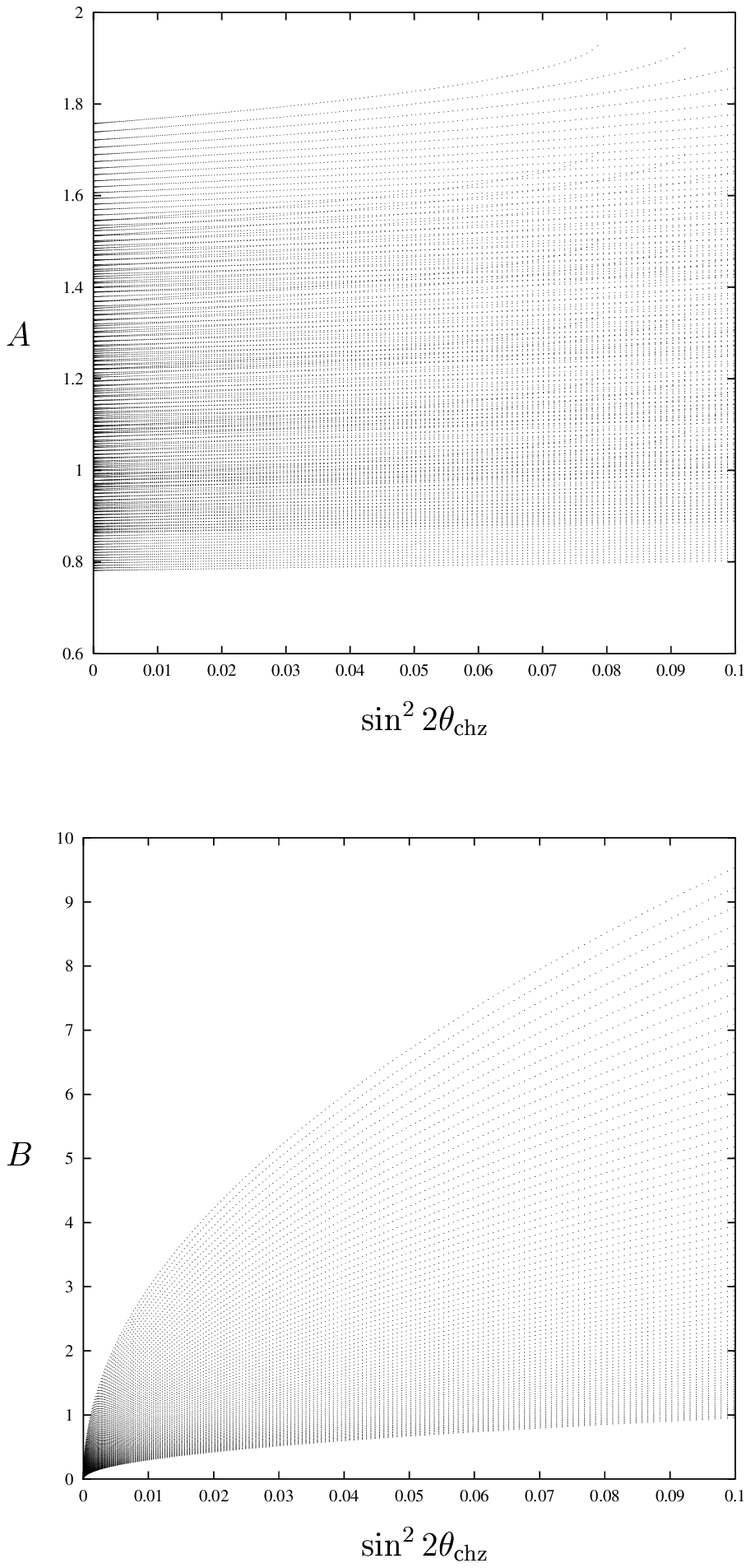,bbllx=3cm,bblly=2cm,bburx=17.5cm,bbury=28cm,%
width=15cm,height=24cm,angle=0,clip=0} 
\vspace{-3cm} 
\caption{Allowed regions of $A$ and $B$ changing with 
$\sin^2 2\theta_{\rm chz}$,
where $0.25 \leq \sin^2\theta_{\rm sun} \leq 0.40$ and 
$\sin^2 2\theta_{\rm atm} > 0.92$ have typically  been input.}
\end{figure}

\end{document}